# The $x$-index: A new citation-distance-based index to measure academic influence

Yun Wan[a], Feng Xiao[a*], Lu Li[b], Zhenghao Zhong[a]

*[a] School of Business Administration, Faculty of Business Administration, Southwestern University of Finance and Economics, PR China*
*[b] Business School, Sichuan University, PR China*

## Abstract

An important issue in the field of academic measurement is how to evaluate academic influence scientifically and comprehensively, which can help government and research organizations better allocate academic resources and recruit researchers. It is generally accepted that using weighted citations to measure academic influence is more reasonable than treating all citations equally. Given the limitations of the existing $c$-index, the first index in bibliometric literature that measures output based on the quantity and quality of received citations, we propose the $x$-index, which assigns weight to each citation according to its distance. By defining collaboration distance and citation distance, we first analyze the properties of the collaboration network and citation distance, then perform theoretical and empirical analyses on $c$-index to reveal its shortcomings, finally, we suggest the $x$-index and conduct experiments and analysis on the $x$-index. Experimental results demonstrate that compared with the $c$-index, $h$-index, and $g$-index, the $x$-index shows a stronger discriminatory power.

---

[*]Corresponding author. E-mail: xiaofeng@swufe.edu.cn

# 1. Introduction

Scholarly impact assessment plays an important role in reward evaluation, funding allocation, promotion, and recruitment decisions (Bai et al., 2017). Moreover, an important issue in the field of academic measurement is how to evaluate academic influence scientifically and comprehensively, which can help government and research organizations better allocate academic resources and recruit researchers. On the one hand, authors hope that their research can be evaluated objectively, fairly, and effectively; on the other hand, readers hope that they can quickly obtain valuable and in-depth scientific research materials through a credible evaluation system.

During the past decades, several bibliometric indicators were proposed to measure the academic impact of scholars, which shows a trend from single measurement to multi-dimensional measurement. Initially, indicators, like the number of citations, the average number of citations of all publications, and the number of highly cited publications, were used to evaluate scholars. However, these metrics are not comprehensive enough as they only consider the numbers of citations. Considering the quality of the article, Hirsch (2005) proposed $h$-index, the number of papers with citation number $\geq h$, which has been widely used for promotions and awards of scholars. Then many indexes were subsequently proposed to make up for the lack of $h$-index (Egghe, 2006; Jin, 2006; Jin et al., 2007; Wu, 2008; Ajiferuke and Wolfram, 2010; Bihari and Tripathi, 2017; Lathabai, 2020). Because of their complicated calculations and the lack of new information added, these indicators have not been promoted. Neither all papers nor all citations should be counted the same way. Scholars generally agree that it is more reasonable to use weighted citations to assess scholars' impact (Cai et al., 2019). Focusing on the issue of how to distinguish the quality of citations, existing studies can be mainly divided into 4 categories:

The first category is based on the collaborative relationship between authors. Some studies (Kosmulski, 2006; Schreiber, 2009; Brown, 2009) removed self-citations in calculating the number of citations directly. Schubert et al. (2006) assigned weights based on the degree of overlap between the authors of the citing article and cited article. Then, Bras-Amorós et al. (2011) suggested the $c$-index, which is the first index in bibliometric literature based on the citation distance: the more distant the citing authors, the higher the quality of a citation.

The second type focuses on the contributions of scholars, which assumes that the contributions of authors of the same article are depending on their rankings and those with higher contributions should be given higher weights. Sekercioglu (2008) defined the k-th ranking as the co-author contributing l/k. Zhang (2009) and Galam (2011) both set a larger weight for the first and corresponding authors.

The third category identifies important citations based on the content of the citation. Wan and Liu (2014) extracted features to identify the importance of citations, such as occurrence number and located section. Valenzuela et al. (2015) defined direct citations and indirect citations, then used 12 related features to identify important citations. Wang et al. (2020) reviewed all these works and divided research perspectives into the categories of citation motivation, citation count, context-based, and metadata.

The fourth category considers scholarly network structure. Westet et al. (2013) developed an author-level Eigenfactor score as a network-based measure of an author's influence. Similarly, Senanayake et al. (2015) and Nykl et al. (2015) used the PageRank algorithm and citation networks to measure and compare the publication records of scientists. Pradhan et al. (2017) used a weighted multi-layered network to rank authors. Further, Franceschet and Colavizza (2018) used citations in a dynamic fashion, allocating ratings by considering the relative position of two authors.

Classifying the importance of citations based on content features or network iteration has a certain theoretical value, however, it is complicated in practical applications. This paper focuses on the $c$-index in the first category. As the first index to consider citation distance, the $c$-index expands the previous consideration from self-citations to greater-distance-based citations. However, the finite distance of citations and the number of citations are not at the same scale, which causes many problems of the $c$-index (details are in Section 4.3). To circumvent these limitations, we propose the $x$-index, which assigns citation weight according to citation distance. We first perform theoretical and empirical analyses on $c$-index, and then conduct experiments and analysis on the $x$-index.

The main contributions of our work are summarized as follows.

(1) To the best of our knowledge, the dataset used in this article is the latest and most complete in the field of computer science in bibliometrics. By processing and analyzing the dataset, we reveal the latest trend in collaboration networks and citation distance.

(2) Through experiments and analysis, the $x$-index has the following advantages:

First, the repeated citations contribute less to $x$-index.

Second, the discriminatory power of $x$-index is stronger than existing indexes.

Third, the $x$-index is difficult to be manipulated and can help identify extreme cases.

Fourth, it can overcome the shortcomings of the $c$-index.

The rest of this article is organized as follows: Section 2 introduces the dataset and some definitions we used in this article. Section 3 provides the property analysis of collaboration networks and citation distance. Analysis of the $c$-index is described in Section 4. Section 5 introduces the $x$-index and performs experiments and analysis. Finally, we conclude the paper in Section 6.

## 2. Dataset and definition

### 2.1 Dataset

We use the dataset DBLP-Citation-network V12 from the Aminer Citation Network Dataset[1] (Tang et al., 2008), which encompasses the field of computer science and contains paper information. In this dataset, some years are unusable because of their incomplete data, so we adopt data from 1966 to 2018 for the analysis. A total of 4,570,521 papers, 4,060,262 scholars, and 41,784,342 citations are included during this period. The dataset, with its long-time span and comprehensive article information, can help us dynamically analyze the trends in this field. As shown in Fig. 1, the yearly numbers of papers and citations both show a clear growth trend.

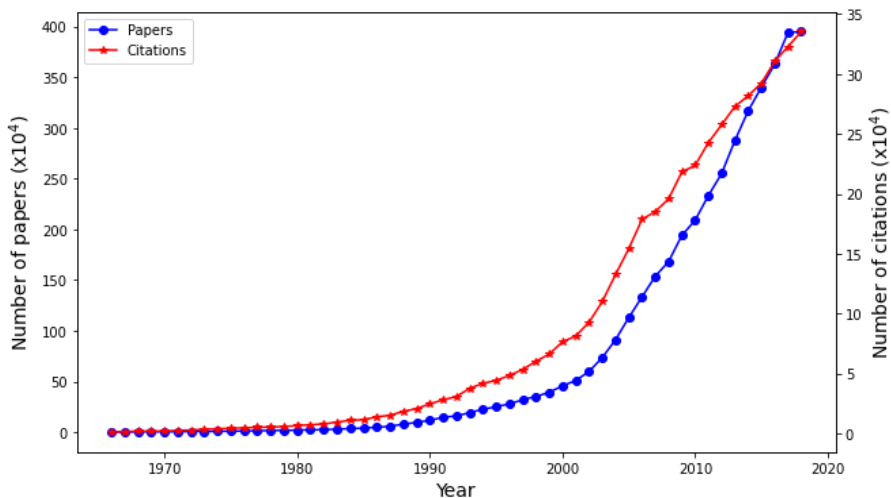

**Fig. 1** The number of papers and citations each year from 1966 to 2018. The blue line with circles represents the number of papers each year; the red line with pentagrams corresponds to the number of citations each year.

---

[1]The citation data are extracted from DBLP, ACM, MAG (Microsoft Academic Graph), and other sources. https://www.aminer.cn/citation

## 2.2 Collaboration network

Two scientists are considered connected if they have authored a paper together (Newman, 2001). Scholars may continue to maintain a good relationship after collaboration. Therefore, it is our view that the time window of collaboration has to be considered. First, we define a collaboration network $C(N, A)$, where $N$ is the set of authors, $A$ is the set of links, $A = \{a_{ij}, i, j \in N, i \neq j\}$, which represents the collaborative relationship between any two nodes. Following Newman (2011), we extract the collaboration relationship in a moving 5-year window, then build the dynamic collaboration network $C^y$, where $y$ represents time. In practical applications, we can determine the update frequency according to needs, such as by month or year. In this work, we update frequency by year because there is no monthly data in the database. Still, we do not believe that this affected our calculation process or results. Edge $a_{ij}$ means that authors $i$ and $j$ have cooperated at least once between $y - 4$ and $y$. For example, the collaboration network $C^{2018}$ contains the cooperation relationship of all papers between 2014 and 2018. For simplicity and computational efficiency, we do not consider the number of collaborations between 2 scholars. Therefore, collaboration network $C$ in this paper is undirected and unweighted.

An example of how to build a dynamic collaboration network is derived as follows: assuming there are 7 papers. Table 1 shows the detailed information of these papers. The first row means that Author 1 and Author 8 published Paper 1 together in 2012. The construction of the collaboration network is shown in Fig. 2 and corresponds to Table 1.

**Table 1** Information of 7 papers

| Papers | Authors | Year |
|---|---|---|
| $p_1$ | 1,8 | 2012 |
| $p_2$ | 1,2,3 | 2013 |
| $p_3$ | 2,3,4 | 2014 |
| $p_4$ | 5,6 | 2015 |
| $p_5$ | 5,6,7 | 2016 |
| $p_6$ | 4,6 | 2017 |
| $p_7$ | 9,2 | 2018 |

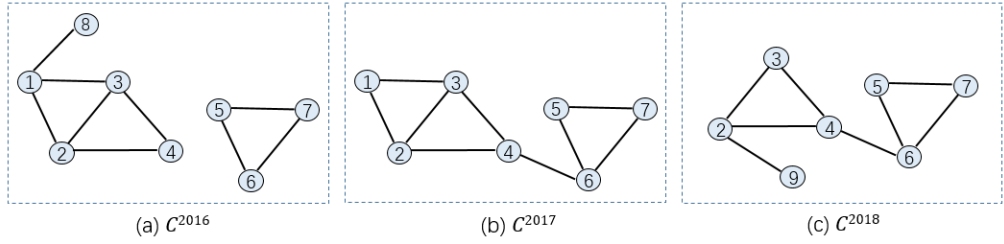

(a) $C^{2016}$  (b) $C^{2017}$  (c) $C^{2018}$

**Fig. 2** Generation of the collaboration network over time: (a) collaboration network of 2016, (b) collaboration network of 2017, (c) collaboration network of 2018.

In this study, we define collaboration distance $\tilde{d}_{ij}$ as the "classical" collaboration distance, which is the distance considered by bibliometric databases such as MathSciNet[2]. Just like the Erdős Number[3], $\tilde{d}_{ij}$ is the shortest path in length between node $i$ and node $j$ in the collaboration network. It equals $\infty$ if there is no reachable path between node $i$ and node $j$, and the scholar is at a distance of 0 from himself or herself.

## 2.3 Citation Network

Corresponding to the citation relationship, we define a citation network $G(M, B)$, where $M$ is the set of papers, $B$ is the set of links. $B = \{b_{ij}, i, j \in M, i \neq j\}$, where a link $b_{ij}$ indicates that paper $i$ is referenced by paper $j$. Therefore, citation network $G$ in this paper is directed and unweighted.

Based on the collaboration distance between scholars, we define the citation distance between articles as follows:

$$d_{ij} = min\{\tilde{d}_{mn} : m \in I, n \in J\} \quad (1)$$

where paper $i$ is cited by paper $j$, $I$ and $J$ are the sets of authors of papers $i$ and $j$, respectively. In the rest of the paper, we always discuss the citation distance of one certain author. Thus, for simplicity, we just ignore the indexes of citing and cited papers, and mark the distance of $m$ citations of the author as $d_1, d_2, \ldots \ldots d_m$.

---

[2] The American Mathematical Society has made an online electronic database MathSciNet, providing all data from 1940 to the present.
[3] The Erdős number is the number of "hops" needed to connect the author of a paper with the prolific late mathematician Paul Erdős. An author's Erdős number is 1 if he has co-authored a paper with Erdős, 2 if he has co-authored a paper with someone who has co-authored a paper with Erdős, for example.

# 3. Property analysis

## 3.1 Properties of collaboration networks

We use 1970, 1980, 1990, 2000, 2010, and 2018 as representatives to reveal the characteristics of collaborative networks in the field of computer science over time. A summary of the basic statistics of these collaboration networks is given in Table 2.

**Table 2** Properties of collaboration networks

| Property | Year | | | | | |
| --- | --- | --- | --- | --- | --- | --- |
| | 1970 | 1980 | 1990 | 2000 | 2010 | 2018 |
| Number of nodes | 5,406 | 20,296 | 84,474 | 302,368 | 1,015,366 | 1,762,080 |
| Number of edges | 4,610 | 22,351 | 116,051 | 593,929 | 2,825,812 | 6,293,688 |
| Average degree | 1.7055 | 2.2025 | 2.7476 | 3.9285 | 5.5661 | 7.1435 |
| Degree assortativity coefficient | 0.7001 | 0.4939 | 0.2919 | 0.4591 | 0.3840 | 0.7263 |
| Average clustering coefficient | 0.3457 | 0.4246 | 0.5134 | 0.6038 | 0.6911 | 0.7458 |
| Number of nodes of 1st connected component | 78 (1.44%) | 1,867 (9.20%) | 20,614 (24.40%) | 151,335 (50.05%) | 651,437 (64.16%) | 1,230,307 (69.82%) |
| Number of edges of 1st connected component | 106 (2.30%) | 3415 (15.28%) | 42,710 (36.80%) | 402,586 (67.78%) | 2,326,689 (82.34%) | 5,464,274 (86.82%) |
| Number of nodes of 2nd connected component | 22 (0.41%) | 61 (0.30%) | 196 (0.23%) | 121 (0.04%) | 200 (0.02%) | 130 (0.01%) |
| Number of edges of 2$^{nd}$ connected component | 25 (0.54%) | 187 (0.84%) | 466 (0.40%) | 816 (0.14%) | 614 (0.02%) | 582 (0.01%) |

(a) Number of nodes and edges. These are basic measures of network size. We observe that the size of the yearly collaboration network increased significantly. This finding is consistent with that of Huang et al. (2008), who found that the number of papers and number of authors increased roughly exponentially over time.

(b) Average degree. According to Bruggeman and Bruggeman (2008), the average degree is simply the average number of edges per node in the graph. It is relatively straightforward to calculate. For collaboration networks, the average degree increases significantly from 1.7055 in 1970 to 7.1435 in 2018, which shows that the collaboration trend strengthened over time.

(c) Degree assortativity coefficient. The assortativity coefficient is the Pearson correlation coefficient of degree between pairs of linked nodes (Newman, 2002). The calculation expression is as follows:

$$r = \sum_{o,k} \frac{ok(e_{ok} - q_o q_k)}{\sigma_q^2} \qquad (2)$$

$$q_k = \sum_o e_{ok} \tag{3}$$

$$\sigma_q^2 = \sum_k k^2 q_k - \left[\sum_k k q_k\right]^2 \tag{4}$$

where $r$ is the degree assortativity coefficient, $o$ and $k$ are the degrees of the 2 vertices of an edge, $e_{ok}$ is the joint-probability distribution of the remaining degrees of the 2 vertices at either end of a randomly chosen edge, and $q_k$ is the edge probability of one vertice, with degree $k$, $\sigma_q^2$ being the variance of the distribution $q_k$. A positive assortativity coefficient value means that nodes tend to connect to the nodes with a similar degree, whereas a negative assortativity coefficient value means that nodes are likely to connect to nodes with very different degrees from their own (Liu et al., 2014). One unanticipated finding is that the coefficient keeps fluctuating. A possible explanation for this might be that with the strengthening of the crossover trend of disciplines and scientific research, scholars are no longer confined to a single field, and thus the mode of cooperation changes. Although the coefficients fluctuate, they are all positive, which indicates that scholars tend to collaborate with scholars who have a similar degree.

(d) Average clustering coefficient. The average clustering coefficient is a global measure of the degree to which nodes in a graph tend to cluster together, which was introduced by Watts and Strogaz (1998). The calculation expression is as follows:

$$\bar{C} = \frac{1}{F} \sum_i \frac{2 t_i}{l_i (l_i - 1)} \tag{5}$$

where $\bar{C}$ is the average clustering coefficient, $F$ is the number of nodes, $l_i$ is the degree of node $i$ in the collaboration network, and $t_i$ is the number of closed triangles containing node $i$. The observed increase of the average clustering coefficient might be due to the transitivity of cooperation of scholars, neighbors of a scholar are more likely to collaborate with each other.

(e) Connected component. In an undirected network, a connected component is a subgraph in which any 2 vertices are interconnected through a path. There are several connected components of a collaboration network. We sort these components in descending order according to the number of nodes and list the first connected component and the second connected component. The number in brackets indicates the percentage of the property of the connected component to that of the total collaboration network. In the yearly collaboration networks, the number of nodes of the largest connected group sharply rises to 69.82% in 2018, containing 86.82% edges. On the contrary, the number of nodes of the second connected component drops to 0.01%, containing only 0.01% edges. This indicates that the vast majority of scientists are connected via collaboration, which makes it feasible and meaningful to calculate the distance.

## 3.2 Properties of citation distance

Given that the calculation of the collaboration network is based on a five-year moving window, we calculate the citation distance from 1970. Among the citations from 1970 to 2018, the largest finite citation distance is 25. We draw the distribution of citation distances as follows:

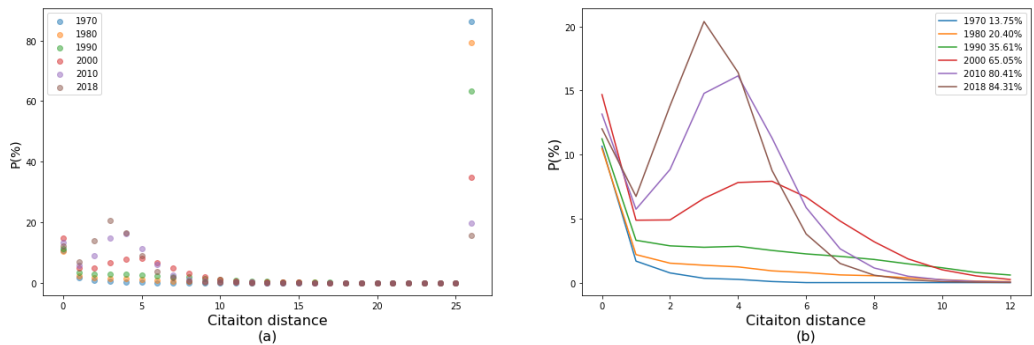

**Fig. 3** (a) is the distribution of citation distance in typical years. The finite distance is between 0 and 25, and the last column of scatter points corresponds to infinite-distance-based citations. (b) is a partition of (a), where citation distance is between 0 and 12. The percentage behind the year represents the proportion of citations with a distance within 12.

Figure 3(a) shows that the proportion of infinite-distance-based citation quickly decreases year over year, from over 80% in 1970 to less than 20% in 2018. Correspondingly, the percentage of close-distance-based citations shows an upward trend. To further reveal the changing law of citation distance, we limit the distance to 12 in Fig. 3(b). From 1970 to 2018, the percentage of self-citations shows a slight rising trend. After year 2000, spikes begin to appear, and the peaks keep shifting to the left until they reach 3 in 2018, which indicates that the citation distance continuously reduces.

# 4. The $c$-index

## 4.1 Definition of the $c$-index

According to Bras-Amorós et al. (2011), given a slope $\alpha$ and a set of citations sorted by decreasing values of distance, the $c$-index of the set can be calculated as

$$c_\alpha = \max\{min(\alpha v, d_v) : v \in (1, \dots, m)\} \qquad (6)$$

where $d_v$ is the distance of the corresponding citation. When evaluating a scholar, we consider all their citations, regardless of which citation corresponds to which paper.

## 4.2 Mathematical meaning of the $c$-index

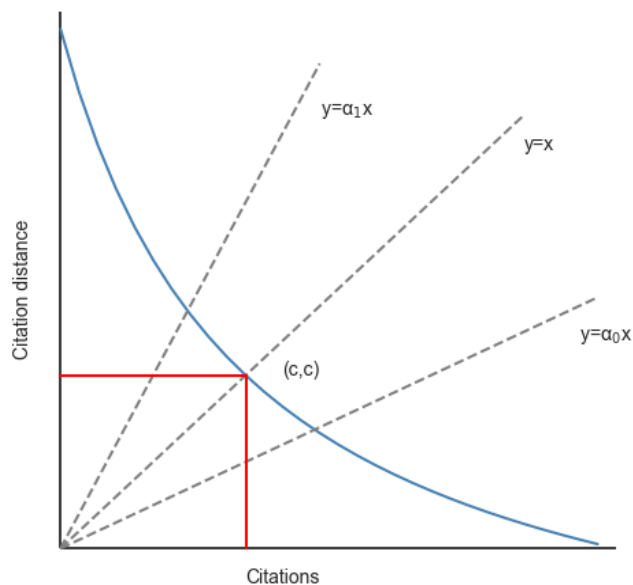

**Fig. 4** $c$-index with different $\alpha$, where $\alpha_1 > 1$ and $\alpha_0 < 1$.

As shown in Fig. 4, arrange all citations in descending order of citation distance, and then correspond their number to the $X$ axis, with each distance corresponding to the $Y$ axis. Draw the citation distance distribution curve of the scholar, the area enclosed by the curve and the coordinate axis equals the total citation distance. Moreover, the $c$-index is the ordinate of the intersection of the straight line and the distribution curve. It can be seen from Fig. 4 that for the same scholar, $c$-index increases with an increase in $\alpha$.

In this study, we set the parameters of $c$-index as recommended by the original paper (Bras-Amorós et.al, 2011), where $\alpha = 1$. However, the value of $\alpha$ does not affect the analysis of the $c$-index, which cannot eliminate problems mentioned later fundamentally. When $\alpha = 1$, the side length of the largest square under the curve with (0,0) as an end point (the oblique line in the figure) is the $c$-index, which is also equal to the coordinate value corresponding to the intersection of the curve and the straight-line $y = x$, which means that the scholar received $c$ citations at a distance of at least $c$ and the rest of citations at a distance of at most $c$. For example, a scholar obtained 10 citations, as shown in Table 3, then his or her $c$-index is 5.

**Table 3** the $c$-index of a scholar

| Rank by distance | Citation distance |
|---|---|
| 1 | 12 |
| 2 | 9 |
| 3 | 7 |
| 4 | 5 |
| **5** | **5** |
| 6 | 4 |
| 7 | 3 |
| 8 | 2 |
| 9 | 1 |
| 10 | 0 |

## 4.3 Properties of the $c$-index

The $c$-index depends on 2 values through analysis: the number of citations and the citation distance. In our dataset, the maximum finite distance is 25, but the maximum number of citations of the scholar is 62,196, which is nearly 2,488 times the maximum finite distance, and it keeps increasing over time. Obviously, the finite distance of citations and the number of citations are not at the same scale. We randomly select a scholar who obtained 107 citations and draw its distribution of citations in Fig. 5.

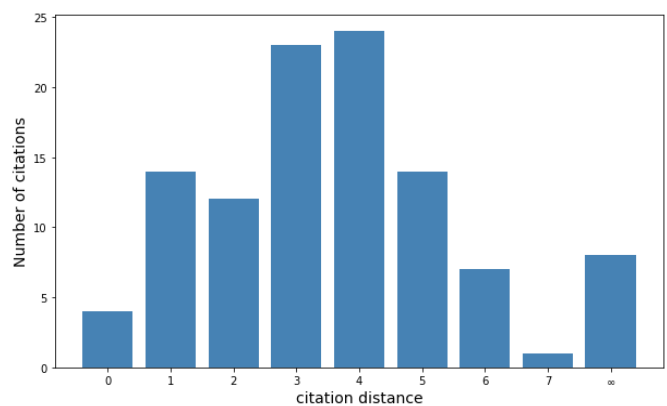

**Fig. 5** The distribution of citations of the scholar.

We count the citations in ascending order based on citation distance. As can be seen from Fig. 5, the distances of most citations are within 7, which is far less than the number of citations. This phenomenon could cause many problems. To further analyze these problems, we mark $D$ as the longest path length of the largest connected component of the collaboration network, $D_f$ as the maximum finite distance of the scholars' citations. It is obvious that $D_f \leq D$. When the citation distance is infinite, we mark it with a symbol, $w$, and define the number of $w$ as $N_w$.

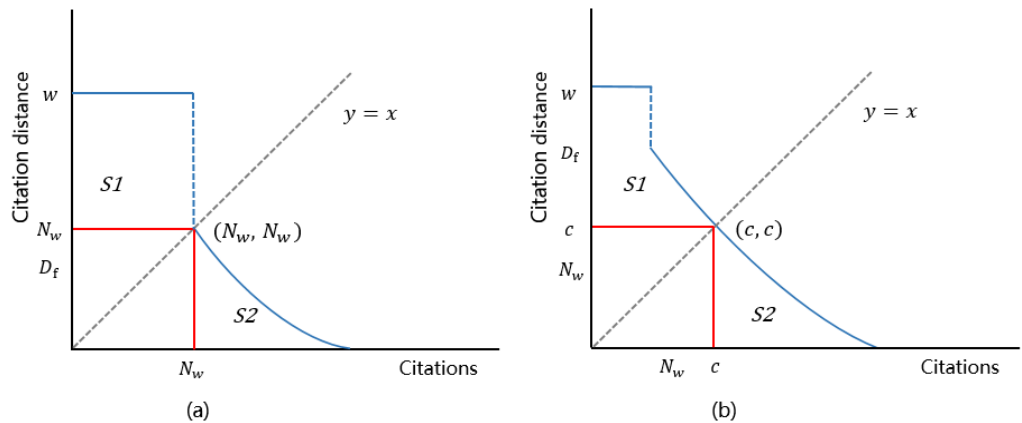

**Fig. 6** Two situations of $N_w$. (a) $N_w > D_f$, (b) $N_w \leq D_f$. The area enclosed by the curve consists of 3 parts, namely, $S1, S2$, and the middle square.

Figure 6 divides the value of $N_w$ into 2 cases: (a) and (b). Combined with these figures, we conclude 3 main disadvantages of the $c$-index.

(1) $c$-index is greatly affected by $w$, which may make the $c$-index meaningless.

When $N_w > D_f$, only $w$ contributes to the $c$-index in Fig. 6(a), where $c$-index is equal to $N_w$. Therefore $c$-index will lose its original meaning. There are 279,825 scholars with citations between 50 and 1000. We divide them into 5 intervals, and calculate the number of scholars, and the number of $c$-index $= N_w$ in each interval. Moreover, the ratio is the proportion of the $c$-index $= N_w$. As can be seen from Table 4, most scholars have the same $c$-index and $N_w$. At the same time, as the number of citations increases, the $c$-index of the scholar is more likely to be equal to $N_w$, which makes the $c$-index lose its original meaning quickly.

**Table 4** Statistics on scholars of $c$-index=$N_w$

| Quantity of citations | Number of scholars | Number of $c$-index=$N_w$ | Ratio |
| --- | --- | --- | --- |
| [50,200) | 200561 | 129183 | 64.41% |
| [200,400) | 46557 | 44143 | 94.81% |
| [400,600) | 17644 | 17519 | 99.29% |
| [600,800) | 9331 | 9315 | 99.82% |
| [800,1000) | 5732 | 5727 | 99.91% |

(2) The discriminatory power of $c$-index is weak.

$S1$ and $S2$ are both important parts of the citation distance, but the $c$-index does not consider these contributions.

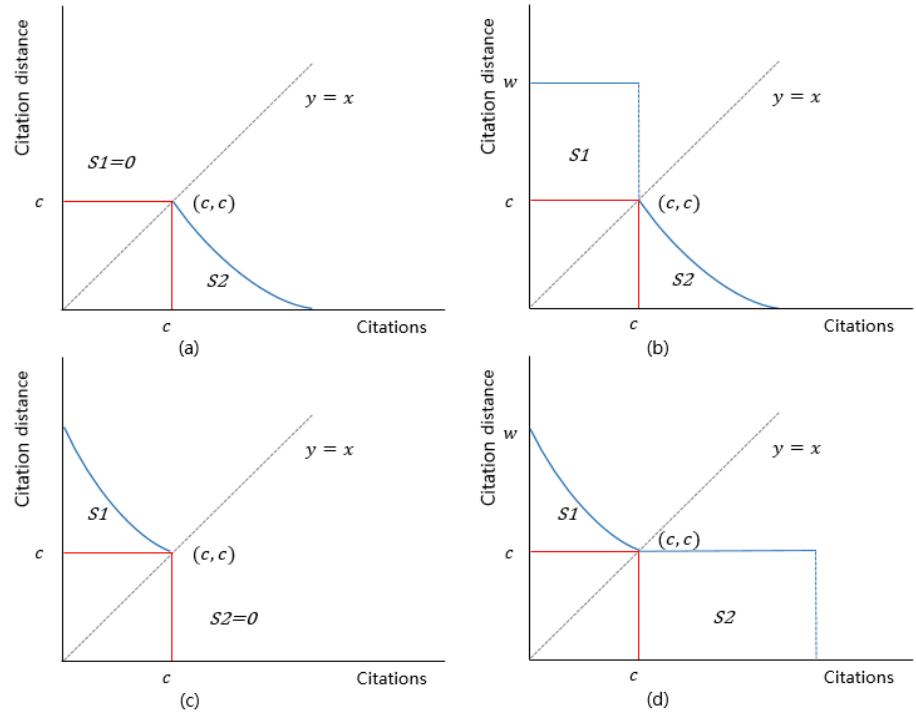

**Fig. 7** The differences of $S1$ and $S2$. (a) is the case where $S1$ takes the minimum value when $N_w \leq D_f$, (b) is the case where $S1$ takes the maximum value when $N_w \leq D_f$, (c) is the case where $S2$ takes the minimum value, (d) is the case where $S2$ takes the maximum value.

As shown in Fig. 7(a) and (b), When $N_w \leq D_f$, if scholars have the same $c$-index, their $S1$ can be 0 or very large, similarly, their $S2$ also can be 0 or infinite. In addition, when $N_w > D_f$ in Fig. 7(c) and (d), if scholars have the same $c$-index, their $S1$ will be the same, but $S2$ may be different. Thus, the difference between $S1$ or $S2$ can be very large, but their $c$-indexes may still be equal, which indicates that the $c$-index has weak discriminatory power. Take $c$-indexes of six scholars in Tables 5 and 6 as examples.

**Table 5** Example I

| Rank by distance | Author | | |
|---|---|---|---|
| | 1 | 2 | 3 |
| 1 | w | w | 5 |
| 2 | w | 10 | 5 |
| 3 | w | 10 | 5 |
| 4 | 12 | 5 | 5 |
| 5 | 5 | 5 | 5 |
| 6 | 4 | 4 | 4 |
| 7 | 3 | 3 | 3 |
| 8 | 2 | 2 | 2 |
| 9 | 1 | 1 | 1 |
| 10 | 0 | 0 | 0 |

**Table 6** Example II

| Rank by distance | Author | | |
|---|---|---|---|
| | 4 | 5 | 6 |
| 1 | 10 | 10 | 10 |
| 2 | 10 | 10 | 10 |
| ... | ... | ... | ... |
| **10** | **10** | **10** | **10** |
| 11 | 10 | 1 | / |
| 12 | 10 | 1 | / |
| ... | ... | ... | ... |
| 30 | 10 | 1 | / |

Author 1,2,……6 are six different scholars arranged in descending order of citation distance. We can see that the distances of the first $c$ citations of Author 1, 2, and 3 vary greatly, however, their c-indexes are the same. Similarly, the numbers and the distances of the citations after $c$ of Author 4, 5, and 6 have a significant difference, but their c-indexes cannot distinguish them.

(3) The requirement for storage space of the $c$-index is high.

The calculation and update of the $c$-index are based on all distances of citations, which have to store the information of all past years. At the same time, the calculations of the $c$-index of the paper and the scholar are independent, so we cannot directly calculate the $c$-index of scholars from the $c$-index of his or her papers.

Based on the above three points, although the $c$-index integrates the quantity and the distance of citations, the information contained is limited, the discriminatory power is weak, and the requirement of storage space is large. To circumvent these limitations, we propose the $x$-index below.

## 5. Introduction of the $x$-index

### 5.1. Definition and calculation framework

In this section, we propose the $x$-index to measure scholars. It assigns weight to each citation according to its distance. The weight is designed as follows:

$$W_d = \begin{cases} f_n(d) & 0 \leq d \leq n \\ 1 & d > n \end{cases} \quad (7)$$

where $d$ represents the distance of each citation, $W_d$ is the weight after mapping, $n$ means that a citation distance of less than or equal to $n$ needs to be mapped. The value of $n$ is directly proportional to the penalty for close citations. The larger $n$ is, the greater the penalty. The value of the citation should increase as the distance increases, and the maximum weight of the citation should be 1, so the mapping function $f_n(d)$ needs to meet the following two criteria:

(1) Monotonic and non-decreasing.

(2) Between 0 and 1.

Many functions can meet such requirements. For simplicity, we choose the piecewise linear function to analyze. It can make the entire mapping function continuous, which is convenient for calculation and comparison.

$$W_d = \begin{cases} \dfrac{d}{n}, & 0 \leq d \leq n \\ 1, & d > n \end{cases} \quad (8)$$

To ensure $W_d$ is meaningful, we set $\dfrac{0}{0} = 1$ here.

After mapping, the $x$-index of scholar $i$ in year $y$ can be calculated by

$$x_i^y = \sum_{t=y_0}^{y} \sum_{d=0}^{\infty} W_d N_{i,d}^t = x_i^{y-1} + \sum_{d=0}^{\infty} W_d N_{i,d}^y = x_i^{y-1} + \Delta x_i^y \quad (9)$$

where $x_i^y$ is the $x$-index of scholar $i$ in year $y$, $N_{i,d}^t$ is the number of citations of $i$ with the distance $d$ in year $t$, $y_0$ is the publication time of the paper, $\Delta x_i^y$ is the increment of $x$-index of $i$ in year $y$. According to Eqs. (9), we only need to store the citation distance of scholar $i$ of the current year.

It is obvious that the value of $n$ reflects the degree of punishment for citation distance. When $n = 0$, the $x$-index reduces to the total number of citations，and the number of citations excluding self-citations is equivalent to the case with $n = 1$. How far the citation should be less weighted is an inconclusive problem. The small-world experiments (Milgram, 1967) are often associated with the phrase "six degrees of separation" (Guare, 1990), which bring the idea that all people on average are 6 or fewer social connections away from each other. Another study (Leskovec & Horvitz, 2008) found the average path length among Microsoft Messenger users to be 6. Based on these experiments, in this study, we assume that citations with a distance greater than 6 are most valuable. Therefore, we set $n$ equals 6 in our experiments.

In this study, we follow the 3 steps below to calculate the $x$-index of the scholar in year $y$:

Step one: build a collaboration network of year $y$ (details can be found in Section 2.2).

Step two: compute the distance of each citation in year $y$ based on the collaboration network.

Step three: update the $x$-index according to Eqs. (9).

## 5.2 Distribution of citation distance

To ensure that the $x$-index is meaningful, we have to confirm that the distributions of citation distances of scholars with similar total numbers of citations are not identical. Otherwise, the $x$-index would have no discriminatory power. We randomly choose 500 scholars with a total number of citations less than 1000 until 2018, and then draw the distributions as follows:

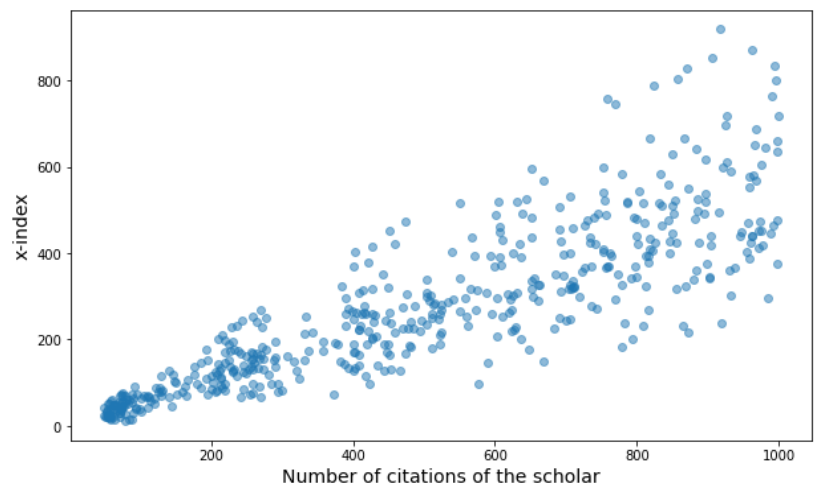

**Fig. 8** $x$-indexes of scholars. Each point represents the $x$-index of a scholar with the same number of citations.

As shown by Fig. 8, the scholars with the same number of citations generally have quite different values of $x$-indexes. Those scholars at the bottom of the scatter chart deserve more attention, since the gaps between their $x$-indexes and the numbers of citations are very large. For example, among those scholars who obtained around 1000 citations, the largest $x$-index is close to 1000, but the smallest $x$-index is just about 200, indicating that most of the citations are based on close distance.

## 5.3 Analysis of the $x$-index

After introducing the definition and calculation process of the $x$-index, in this section, we further investigate the properties of this index.

By definition, the value of $x$-index is between 0 and the total number of citations. Considering both the number of citations and the quality of citations, the $x$-index offers the following key advantages:

i) The repeated citations contribute less to $x$-index.

Here, we ignore the citation direction and count the number of citations between two scholars. For example, if scholar $a$ cited scholar $b$ once, and scholar $b$ cited scholar $a$ once, then we count the number of citations between $a$ and $b$ as 2. If this number is greater than 1, we regard the rest of the citations as repeated citations. We count the numbers of citations between each pair of scholars from year 2014 to 2018 and calculate the collaboration distance between them. Then, we draw the heatmap shown in Fig. 10. As shown by Fig. 10, the highlighted area in each row keeps moving to the left from bottom to top, and getting brighter. This means that the shorter the collaboration distance between scholars, the more the number of repeated citations. Since under the calculation of $x$-index, a closer collaboration distance will be assigned a smaller weight, by definition the repeated citations will contribute less to the value of $x$-index.

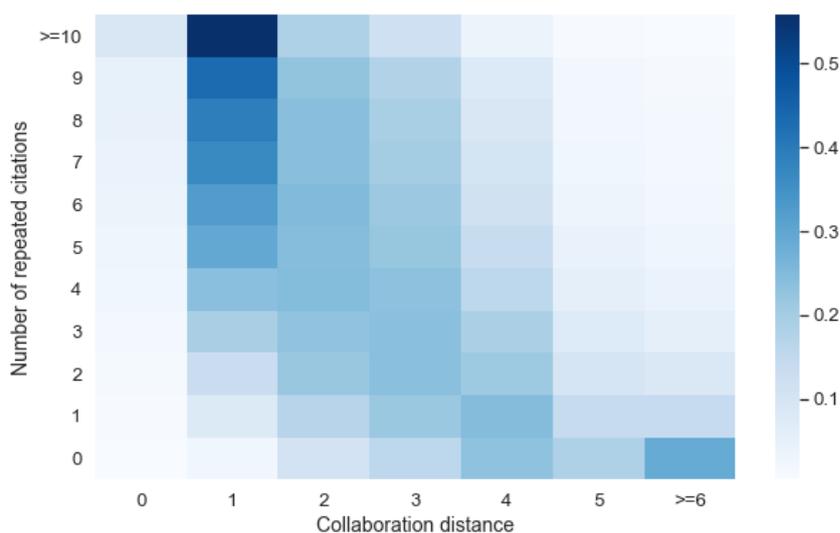

Fig. 10 Heatmap of the number of repeated citations and collaboration distance

ii) The discriminatory power of $x$-index is stronger than existing indexes.

We compare the $x$-index of scholars with the $c$-index, $h$-index, and $g$-index in turn. To ensure that our experimental results are not affected by other factors, we first fix the total number of citations to be between 190 and 210 (for most of the scholars are within this range); then we fix the values of other indexes when compare $x$-index with one certain index. Finally, we randomly select 20 scholars that satisfy the above conditions to compare their indexes and rankings. We calculate the standard deviation of each index, and mark any pair of scholars as 'Close' if the difference between their values of the index is within 0.1 times the standard deviation. Then, the comparison results of scholars between $x$-index and any other index can be divided into the following four situations in Table 7.

Table 7 Four situations of results

| One certain index | $x$-index |
|---|---|
| Close | Close |
| Close | Not Close |
| Not Close | Close |
| Not Close | Not Close |

The comparisons between $x$-index and other indexes are as follows:

Experiment I: Comparison between the $x$-index and the $c$-index.

Table 8 provides the index values and rankings of the 20 scholars. Table 9 shows the pairs of scholars under the four situations in Experiment I.

Table 8 Indexes of the 20 scholars in Experiment I

| ID | $Q$ | $h$-index | $g$-index | $N_w$ | $c$-index | $r^c$ | $x$-index | $r^x$ |
|---|---|---|---|---|---|---|---|---|
| 1 | 197 | 8 | 13 | 120 | 120 | 1 | 170.33 | 4 |
| 2 | 208 | 8 | 13 | 103 | 103 | 2 | 149.00 | 10 |
| 3 | 191 | 8 | 13 | 91 | 91 | 3 | 172.33 | 2 |
| 4 | 192 | 8 | 13 | 85 | 85 | 4 | 104.67 | 16 |
| 5 | 199 | 8 | 13 | 80 | 80 | 5 | 152.00 | 9 |
| 6 | 200 | 8 | 13 | 77 | 77 | 6 | 183.00 | 1 |
| 7 | 202 | 8 | 13 | 72 | 72 | 7 | 164.00 | 6 |
| 8 | 206 | 8 | 13 | 70 | 70 | 8 | 167.00 | 5 |
| 9 | 191 | 8 | 13 | 69 | 69 | 9 | 163.50 | 7 |
| 10 | 191 | 8 | 13 | 64 | 64 | 10 | 103.50 | 17 |
| 11 | 192 | 8 | 13 | 62 | 62 | 11 | 138.67 | 12 |
| 12 | 198 | 8 | 13 | 60 | 60 | 12 | 121.67 | 13 |
| 13 | 207 | 8 | 13 | 55 | 55 | 13 | 99.00 | 18 |
| 14 | 191 | 8 | 13 | 50 | 50 | 14 | 157.50 | 8 |
| 15 | 192 | 8 | 13 | 46 | 46 | 15 | 171.17 | 3 |
| 16 | 195 | 8 | 13 | 40 | 40 | 16 | 147.33 | 11 |
| 17 | 195 | 8 | 13 | 35 | 35 | 17 | 92.00 | 19 |
| 18 | 198 | 8 | 13 | 30 | 30 | 18 | 79.67 | 20 |
| 19 | 195 | 8 | 13 | 25 | 25 | 19 | 115.67 | 14 |

| 20 | 199 | 8 | 13 | 20 | 20 | 20 | 113.33 | 15 |
| s | | | | | 25.49 | | 30.83 | |

Here, $Q$ represents the total number of citations. $r^c$ is the ranking based on $c$-index, and $r^x$ is the ranking based on $x$-index. If the index values are equal, we rank them according to $Q$. $s$ is the standard deviation of the values of indexes.

Table 9 Selection of the four pairs of scholars in the four situations in Experiment I

| $c$-index | $x$-index | ID |
| --- | --- | --- |
| Close | Close | 7,8 |
| Close | Not Close | 10,11 |
| Not Close | Close | 3,15 |
| Not Close | Not Close | 4,6 |

To analyze these four pairs of scholars in Table 9, we list the number of citations at different distances of these scholars, as shown in Table 10.

Table 10 Citation distances in Experiment I

| citation distance | ID | | | | | | | |
| --- | --- | --- | --- | --- | --- | --- | --- | --- |
| | 7 | 8 | 10 | 11 | 3 | 15 | 4 | 6 |
| 0 | 20 | 26 | 24 | 19 | 11 | 9 | 59 | 13 |
| 1 | 1 | 0 | 29 | 11 | 0 | 4 | 28 | 1 |
| 2 | 4 | 1 | 41 | 16 | 1 | 2 | 6 | 0 |
| 3 | 6 | 2 | 16 | 11 | 2 | 7 | 1 | 1 |
| 4 | 19 | 25 | 10 | 19 | 10 | 5 | 0 | 5 |
| 5 | 31 | 18 | 4 | 16 | 16 | 12 | 3 | 6 |
| ⩾6 | 121 | 134 | 67 | 100 | 151 | 153 | 95 | 174 |

From the above tables, we have the following observations:

First, scholars 7 and 8 have a similar $c$-index and $x$-index, the distributions of citation distance are also similar.

Second, although scholars 10 and 11 have a similar $c$-index, their $x$-indexes are quite different, which is mainly because scholar 10 has more close citations than scholar 11. Similarly, too many close citations cause scholar 4's ranking to drop from $4th$ in the $c$-index to $16th$ in the $x$-index.

Third, scholars 3 and 15 have a similar $x$-index, but the $c$-index of scholar 3 is two times of that of Scholar 2, which is because their $c$-indexes are only affected by $N_w$.

(2) Experiment II: Comparison between the $x$-index and the $h$-index.

Table 11 provides the indexes and rankings of the 20 scholars. Table 12 shows the selection of four pairs of scholars under four situations in Experiment II, and Table 13 list the numbers of citations at different distances of these scholars.

**Table 11** Indexes of the 20 scholars in Experiment II

| ID | Q | g-index | c-index | $N_w$ | h-index | $r^h$ | x-index | $r^x$ |
|---|---|---|---|---|---|---|---|---|
| 1 | 204 | 14 | 10 | 10 | 10 | 1 | 65.00 | 16 |
| 2 | 200 | 14 | 10 | 10 | 10 | 2 | 78.67 | 14 |
| 3 | 207 | 14 | 10 | 10 | 9 | 3 | 88.00 | 9 |
| 4 | 199 | 14 | 10 | 10 | 9 | 4 | 45.67 | 20 |
| 5 | 208 | 14 | 10 | 10 | 8 | 5 | 150.17 | 1 |
| 6 | 204 | 14 | 10 | 10 | 8 | 6 | 99.83 | 8 |
| 7 | 198 | 14 | 10 | 10 | 8 | 7 | 62.67 | 18 |
| 8 | 202 | 14 | 10 | 10 | 7 | 8 | 106.50 | 3 |
| 9 | 201 | 14 | 10 | 10 | 7 | 9 | 86.67 | 11 |
| 10 | 199 | 14 | 10 | 10 | 7 | 10 | 106.00 | 4 |
| 11 | 204 | 14 | 10 | 10 | 6 | 11 | 55.33 | 19 |
| 12 | 199 | 14 | 10 | 10 | 6 | 12 | 106.00 | 5 |
| 13 | 204 | 14 | 10 | 10 | 5 | 13 | 65.67 | 15 |
| 14 | 196 | 14 | 10 | 10 | 5 | 14 | 80.33 | 13 |
| 15 | 210 | 14 | 10 | 10 | 4 | 15 | 113.17 | 2 |
| 16 | 199 | 14 | 10 | 10 | 4 | 16 | 63.00 | 17 |
| 17 | 197 | 14 | 10 | 10 | 4 | 17 | 102.67 | 6 |
| 18 | 205 | 14 | 10 | 10 | 3 | 18 | 100.00 | 7 |
| 19 | 198 | 14 | 10 | 10 | 3 | 19 | 87.50 | 10 |
| 20 | 196 | 14 | 10 | 10 | 2 | 20 | 81.00 | 12 |
| s | | | | | 2.36 | | 23.80 | |

$r^h$ is the result of ranking according to $h$-index; when the $h$-index is equal, we sort the scholars according to $Q$.

**Table 12** Selection of four pairs of scholars under four situations in Experiment II

| h-index | x-index | ID |
|---|---|---|
| Close | Close | 8,10 |
| Close | Not Close | 3,4 |
| Not Close | Close | 6,18 |
| Not Close | Not Close | 1,12 |

**Table 13** Citation distances in Experiment II

| citation distance | ID | | | | | | | |
|---|---|---|---|---|---|---|---|---|
| | 8 | 10 | 3 | 4 | 6 | 18 | 1 | 12 |
| 0 | 30 | 27 | 39 | 105 | 28 | 21 | 52 | 15 |
| 1 | 10 | 12 | 35 | 23 | 11 | 10 | 50 | 9 |
| 2 | 24 | 16 | 28 | 18 | 28 | 25 | 39 | 27 |
| 3 | 53 | 49 | 35 | 27 | 65 | 91 | 26 | 67 |
| 4 | 32 | 50 | 36 | 8 | 38 | 35 | 17 | 48 |
| 5 | 24 | 25 | 16 | 6 | 19 | 11 | 4 | 18 |
| ⩾6 | 29 | 20 | 18 | 12 | 15 | 12 | 16 | 15 |

From the tables above, we can observe that:

First, when scholars' citations are similar in the distributions of both papers and distances, their $h$-indexes and $x$-indexes are also similar, like scholars 8 and 10.

Second, although scholars 3 and 4 have the same $h$-index, $x$-index of scholar 3 is much larger than that of scholar 4, which is mainly because over half of scholar 4's citations are self-citations. Similarly, too many close citations cause scholar 1's ranking to drop from $1^{th}$ in the $h$-index to $16^{th}$ in the $x$-index.

Third, scholars 6 and 18 have similar $x$-indexes, but scholar 6's $h$-index is much higher than that of scholar 18, which is mainly because the $h$-index is only affected by the number of papers and the distribution of papers' citations.

(3) Experiment III: Comparison between the $x$-index and the $g$-index.

Table 14 provides the indexes and rankings of the 20 scholars. Table 15 shows the selection of four pairs of scholars under four situations Experiment III, respectively, and Table 16 lists the number of citations at different distances of these scholars.

**Table 14** Indexes of the 20 scholars in Experiment III

| ID | $Q$ | $h$-index | $c$-index | $N_w$ | $g$-index | $r^g$ | $x$-index | $r^x$ |
|---|---|---|---|---|---|---|---|---|
| 1 | 205 | 8 | 9 | 9 | 14 | 1 | 29.33 | 20 |
| 2 | 203 | 8 | 9 | 9 | 14 | 2 | 88.83 | 5 |
| 3 | 203 | 8 | 9 | 9 | 14 | 3 | 119.33 | 2 |
| 4 | 202 | 8 | 9 | 9 | 14 | 4 | 103.33 | 4 |
| 5 | 205 | 8 | 9 | 9 | 13 | 5 | 54.50 | 18 |
| 6 | 201 | 8 | 9 | 8 | 13 | 6 | 73.83 | 10 |
| 7 | 197 | 8 | 9 | 9 | 13 | 7 | 115.50 | 3 |
| 8 | 194 | 8 | 9 | 9 | 13 | 8 | 55.83 | 16 |
| 9 | 190 | 8 | 9 | 9 | 13 | 9 | 73.17 | 11 |
| 10 | 207 | 8 | 9 | 9 | 12 | 10 | 82.33 | 8 |
| 11 | 199 | 8 | 9 | 9 | 12 | 11 | 68.33 | 12 |
| 12 | 197 | 8 | 9 | 9 | 12 | 12 | 83.67 | 7 |
| 13 | 197 | 8 | 9 | 9 | 12 | 13 | 66.33 | 13 |
| 14 | 204 | 8 | 9 | 9 | 11 | 14 | 125.83 | 1 |
| 15 | 202 | 8 | 9 | 9 | 11 | 15 | 59.67 | 15 |
| 16 | 199 | 8 | 9 | 7 | 11 | 16 | 81.00 | 9 |
| 17 | 196 | 8 | 9 | 9 | 11 | 17 | 88.83 | 6 |
| 18 | 196 | 8 | 9 | 9 | 10 | 18 | 65.17 | 14 |
| 19 | 210 | 8 | 9 | 9 | 9 | 19 | 44.33 | 19 |
| 20 | 190 | 8 | 9 | 7 | 9 | 20 | 55.50 | 17 |
| s | | | | | 1.53 | | 24.63 | |

$r^g$ is the result of ranking according to $g$-index. When the $g$-index is equal, we sort the scholars according to $Q$.

**Table 15** Scholar pairs of four situations in Experiment III

| $g$-index | $x$-index | ID |
|---|---|---|
| Close | Close | 10,12 |
| Close | Not Close | 1,3 |
| Not Close | Close | 8,20 |
| Not Close | Not Close | 5,14 |

**Table 16** Citation distances in Experiment III

| citation distance | ID | | | | | | | |
|---|---|---|---|---|---|---|---|---|
| | 10 | 12 | 1 | 3 | 8 | 20 | 5 | 14 |
| 0 | 86 | 49 | 130 | 36 | 95 | 124 | 117 | 65 |
| 1 | 1 | 6 | 41 | 3 | 25 | 4 | 8 | 26 |
| 2 | 6 | 29 | 8 | 2 | 10 | 0 | 12 | 24 |
| 3 | 33 | 47 | 9 | 32 | 15 | 2 | 22 | 41 |
| 4 | 40 | 44 | 5 | 64 | 18 | 13 | 21 | 17 |
| 5 | 24 | 11 | 0 | 39 | 13 | 11 | 5 | 11 |
| ⩾6 | 17 | 11 | 12 | 27 | 18 | 36 | 20 | 13 |

The $g$-index is an improvement of $h$-index, but it is highly correlated with $h$-index. Thus, the analyses of experiment II and experiment III are similar:

First, scholars 10 and 12 have a similar $g$-index and $x$-index. Although scholar 10 has a lot of self-citations, his or her distant citations make up for these self-citations.

Second, although scholars 1 and 3 have the same $g$-index, scholar 1's ranking drops from the first in the $g$-index to the last in the $x$-index, which is mainly because scholar 1 has too many self-citations. Similarly, too many self-citations lead to a low $x$-index for scholar 5.

Third, scholars 8 and 20 have a similar $x$-index, but scholar 8's $g$-index is higher than that of scholar 20, since the $g$-index is affected by the distribution of papers' citations.

Through the above three experiments, we can find that compared with other indexes, the $x$-index is more powerful in distinguishing close citations.

iii) The $x$-index is difficult to be manipulated and can help identify extreme cases.

If scholars collaborated before trying to enhance the index by an improper citation relationship for intentional mutual citation, this would quickly increase the scholar's $h$-index and $g$-index, but it would not have much impact on the $x$-index. The closer the citation distance, the smaller the impact on $x$-index. As such, if one wants to manipulate the $x$-index, he must increase the number of distant citations, which is difficult, if not impossible. At the same time, $x$-index can help us identify those abnormal manipulation behaviors. For example, in Table 15, $x$-index can help identify the extreme cases as scholars 1 and 5.

iv) $x$-index can overcome the shortcomings of the $c$-index mentioned in Section 4.3.

In addition to a better discriminatory power of the $x$-index mentioned in the previous experiments, the $x$-index is not like the $h$-index or $c$-index, which needs to store the information of all past years; instead, it can be directly updated based on the old $x$-index of last year.

## 6. Discussion and conclusions

To circumvent the limitations of $c$-index, we proposed the $x$-index, which assigns citation weight according to citation distance, and analyzed its properties. It should be noted that the $x$-index also can be used for the evaluation of articles and groups, and the calculation method is the same as that of scholars. In addition, since the measurement is additive, one scholar's $x$-index can be calculated directly by accumulating the $x$-indexes of all the papers he/her has published. We proved that the discriminatory power of $x$-index is stronger than several widely used existing indexes. And it is difficult to be manipulated while easy to be updated.

It should be pointed out that this research could be extended in the future in at least the following two aspects: 1) in terms of data, this study only used the dataset in the area of computer science from 1966 to 2018. Taking the differences of disciplines into account, future experiments should cover more academic areas for further verification. 2) we do not consider other relevant factors, such as the content of citation, the influence of the journal, and the different contributions of the coauthors. Thus, this index should be combined with other indexes to measure the influence of scholars more comprehensively.

## Declarations

### Acknowledgments

This work is supported by National Science Fund for Distinguished Young Scholars (Project No.72025104). And we thank Prof. Bintong Chen for his valuable advice.

## Conflicts of interest

The authors have no relevant financial or non-financial interests to disclose.

## Authors' contribution

The initial idea was first proposed by Feng Xiao. Material preparation, data collection and analysis were performed by Yun Wan, Feng Xiao and Lu Li. The first draft of the manuscript was written by Yun Wan and all authors commented on previous versions of the manuscript. All authors read and approved the final manuscript.